\documentclass[prd,twocolumn,showpacs,superscriptaddress,nofootinbib]{revtex4-2}

\usepackage{amsfonts}
\usepackage{amsmath}
\usepackage{amssymb}
\usepackage{amsthm}
\usepackage{bm}
\usepackage{dcolumn}
\usepackage{epsfig}
\usepackage{graphicx}
\usepackage{graphics}
\usepackage[latin1]{inputenc}
\usepackage{latexsym}
\usepackage{rotating}
\usepackage{hyperref}
\usepackage{ulem}

\usepackage[dvipsnames, usenames]{xcolor}

\newcommand\be{\begin{equation}}
\newcommand\ba{\begin{eqnarray}}
\newcommand\ee{\end{equation}}
\newcommand\ea{\end{eqnarray}}

\newcommand{\pont}{{\,^\ast\!}R\,R}
\newcommand{\static}{\mbox{\tiny st}}

\newcommand{\EH}{{\mbox{\tiny EH}}}
\newcommand{\K}{{\mbox{\tiny K}}}
\newcommand{\ergo}{{\mbox{\tiny ergo}}}

\newcommand{\Brown}{Brown Theoretical Physics Center and Department of Physics, Brown University, Providence, Rhode Island, 02903}

\begin{document}
\title{The Chern-Simons Caps for Rotating Black Holes}

\author{Stephon Alexander}
\affiliation{\Brown}
\affiliation{CCA, Flatiron Institute, 162 5th Ave, NY, New York}

\author{Gregory Gabadadze}
\affiliation{Center for Cosmology and Particle Physics, Department of Physics, New York University, New York, New York 10003}
\author{Leah Jenks}
\affiliation{\Brown}
\author{Nicol\'as Yunes}
\affiliation{Illinois Center for Advanced Studies of the Universe, Department of Physics, University of Illinois at Urbana-Champaign}

\date{\today}

\preprint{IGC-07/11-1}

%%%%%%%%%%%%%%%%%%%%%%%%%%%%%%%%%%%%%%%%%%%%%%%%%%%%%%%%%%%%%%%%%%%%%%%%%%%%%%
\begin{abstract}
  
We study the dynamical Chern-Simons gravity as an effective quantum field theory, 
and discuss a broad range of its parameter space where the theory is valid.  Within that validity range,  
we show that slowly rotating black holes acquire  novel geometric structures due to the gravitational dynamical  
Chern-Simons term. In particular, the rotating black hole solutions get endowed with two, 
cap-like domains,  emanating from the north and south poles in the standard 
Boyer-Lindquist coordinates. The domains extend  out to a distance that  is approximately a few percent of the black 
hole's  size.  The cap-like domains have an unusual  equation of state, pointing to non-standard 
dynamics within the caps.  In particular, the  focusing condition for geodesics is violated in
those domains. This in turn implies that the Hawking-Penrose singularity theorem cannot be straightforwardly 
applied to hypothetical probe matter placed within the Chern-Simons caps. 
\end{abstract}

\pacs{04.20.Cv,04.70.Bw,04.20.Jb,04.30.-w}

\maketitle

%%%%%%%%%%%%%%%%%%%%%%%%%%%%%%%%%%%%%%%%%%%%%%%%%%%%%%%%%%%%%%%%%%%%%%%%%%%%%%
\section{Introduction and outline}
\label{intro}

Dynamical Chern-Simons (dCS) gravity~\cite{Jackiw:2003pm, Alexander:2009tp} modifies the Einstein-Hilbert action with the addition of a parity-violating Chern-Simons form coupled to a derivative of a pseudo-scalar field.  
dCS gravity is not an arbitrary extension of general relativity (GR), but rather has physical roots in particle physics~\cite{ALVAREZGAUME1984269} and string theory~\cite{Polchinski:1998rr, PhysRevLett.96.081301, Alexander:2004xd}. dCS gravity naturally emerges as an anomaly-canceling term through the Green-Schwarz mechanism~\cite{Green:1987mn}.  It is often the case that  the scale  at which  the dCS term becomes 
significant is inseparable  from the scale where other higher dimensional terms become significant, too. 

In this work we regard the stand-alone dCS action as a low energy  effective field theory.   
In order  for us to be able to keep the dCS term in the action, while neglecting 
an infinite number of the Planck mass suppressed 
terms, we require that the  scale where the CS term becomes relevant is much smaller than the 
Planck mass. We specify  a  field theory mechanism that  can lead to such a hierarchy of scales without 
unnatural fine tunings, and discuss a broad range of the parameter space  where our approach is justified. 
The above discussions are presented in Section 2.

Within the range of validity of the effective field theory we then look for certain novel 
characteristics of rotating black hole solutions in Section 3.
Previous work has shown that dCS gravity admits some solutions of GR 
without any obstruction, while predicting modifications to GR solutions that lack a sufficiently high degree of symmetry~\cite{jackiw, CAMPBELL1991778, Guarrera:2007tu, Grumiller:2007rv}. 
In particular, the dCS theory supplies an additional term in the Einstein equation, which 
can be thought of as a new "stress-energy tensor" in the right-hand side of the equation. 
Since this new term descends from a quantum anomaly, it does not have to obey 
the classical positivity conditions. The latter feature manifests  itself in the properties 
of the solutions of the theory that  have a nonzero new ``stress-energy tensor.''  Such are the rotating 
black hole solutions carrying dCS pseudoscalar hair \cite{Yunes:2009hc,Yagi:2012ya,Maselli:2017kic}. 

We show, in Section 3,  that the rotating black holes possess   a novel geometric 
structure due to the dCS term. In particular, the solutions get endowed with two, 
cap-like domains,  emanating from the north and south poles of the black holes expressed 
in the standard Boyer-Lindquist coordinates. These new domains, referred to here as ``CS caps,'' extend  
out to a distance that  is approximately a few percent of the black hole's  size.  The CS caps have an unusual  
equation of state, which leads to the violation of the focusing condition for geodesics.

While the CS caps of the rotating black holes  may have some interesting observational consequences
(to be investigated in subsequent works), the fact that they violate the focusing condition calls 
for rethinking of the Hawking-Penrose (HP) theorem ~\cite{Penrose:1964wq, Hawking:1969sw}, 
as we discuss in Section 4. According to our findings, the HP theorem cannot be applied to geodesics of external probe 
matter placed within the CS caps, where the focusing condition is violated. While this fact itself
says nothing about the singularity (non)formation for  external matter in those domains, 
it nevertheless represents an existence proof of a stable spatial domains where the main 
condition of the HP theorem is not fulfilled.

%%%%%%%%%%%%%%%%%%%%%%%%%%%%

%%%%%%%%%%%%%%%%%%%%%%%%%%%%%%%%%%%%%%%%%%%%%%%%%%%%%%%%%%%%%%%%%%%%%%%5

We use the following conventions in this paper: we work exclusively in
four spacetime dimensions with signature
$(-,+,+,+)$, with Latin letters $(a,b,\ldots,h)$
ranging over all spacetime coordinates; round and
square brackets around indices denote symmetrization and
anti-symmetrization respectively, namely $T_{(ab)}:=\frac12
(T_{ab}+T_{ba})$ and $T_{[ab]}:=\frac12 (T_{ab}-T_{ba})$; partial
derivatives are sometimes denoted by commas, e.g.~$\partial
\theta/\partial r=\partial_r\theta=\theta_{,r}$. The Einstein
summation convention is employed unless otherwise specified.

%%%%%%%%%%%%%%%%%%%%%%%%%%%%%%%%%%%%%%%%
\section{DCS as an Effective Field Theory}
\label{EFT}

Let us begin by defining the action of dCS gravity~\cite{jackiw}:
\be
\label{CSsigma}
S =  \int d^4x \sqrt{-g} \left[\kappa R + \frac{\sigma }{4\mu } \; \pont
- \frac{1}{2} \left(\nabla_a \sigma  \right) \left(\nabla^a \sigma \right) \right]\,, 
\ee
where $\kappa = (16 \pi G_N)^{-1}$, $g$ is the determinant of the metric, the integral extends over all spacetime, $R$ is the Ricci scalar, and the pseudo-scalar field $\sigma$  couples to the Pontryagin invariant $\pont$,  defined as follows
\be
\label{pontryagindef}
\pont:={\,^\ast\!}R^a{}_b{}^{cd} R^b{}_{acd}\,,
\ee
where the dual Riemann-tensor is given by
\be
\label{Rdual}
{^\ast}R^a{}_b{}^{cd}:=\frac12 \epsilon^{cdef}R^a{}_{bef}\,,
\ee
with $\epsilon^{cdef}$ the 4-dimensional Levi-Civita tensor\footnote{We prefer to work with tensors rather than with tensor densities in this paper, so some expressions might appear to differ by factors of $\sqrt{-g}$ from \cite{Jackiw:2003pm}.}. The Pontryagin term [Eq.~\eqref{pontryagindef}] can be expressed as the divergence
\be
\nabla_a K^a = \frac14 \pont 
\label{eq:curr1}
\ee
of the Chern-Simons topological current ($\Gamma$ is the Christoffel connection), 
\be
K^a :=\epsilon^{abcd}\left(\Gamma^n{}_{bm}\partial_c\Gamma^m{}_{dn}+\frac23\Gamma^n{}_{bm}\Gamma^m{}_{cl}\Gamma^l{}_{dn}\right)\,.
\label{eq:curr2}   
\ee
Hence, the interaction term of $\sigma$  can be rewritten, up to a total derivative, as follows  
\be
\label{CSsigmaCS}
 - \frac{ 1 }{\mu }  \int d^4x \sqrt{-g} \, g_{ab} \, \partial^a \sigma \, K^b\,. 
\ee
We will use interchangeably the term in Eq.~(\ref {CSsigmaCS})  with  its counterpart  in Eq.~(\ref {CSsigma}),
keeping in mind that there is a total derivative distinguishing the two. Note that the action in Eq.~(\ref {CSsigmaCS})
makes the shift symmetry of the  pseudoscalar field manifest, $\sigma\to\sigma +const$.
In this formulation, the shift symmetry current on a non-interacting theory, $\partial_b \sigma$, couples
to the Chern-Simons current, $K^b$, thus the name ``Chern-Simons modified gravity''\footnote{If $\nabla_a K^a$ is converted into $1/\sqrt{g} \, \partial_a (\sqrt{g} K^a)$ the results (2.4) and (2.5) of \cite{Jackiw:2003pm} are recovered.}. 

%The term that couples the pseudoscalar field to the Pontryagin density is a higher dimensional operator of an effective quantum field theory described by Eq.~(\ref{CSaction}).

General Relativity is not a renormalizable theory, and hence, the action in Eq.~(\ref {CSsigma})  could only be 
part of an effective field theory that contains an infinite number of  higher curvature terms, proportional to $R^2, R^3$ and so on,
and their derivatives, suppressed by the respective 
powers of the Planck mass, $M_P=1/ \sqrt{G_N}$. In order for the  higher dimensional terms 
to be negligible  as compared  to the  term   $\sigma \pont /({4\mu })$ kept in Eq.~(\ref {CSsigma}), 
we should require
\be
\mu << M_P\,.
\ee
Since the term ${\sigma \pont }/({4\mu })$ is not renormalizable either, its presence  would then imply 
some new physics at the scale $\mu <<M_P$.  The new physics would  generate the term
${\sigma \pont }/({4\mu })$ at low energies, $E<<\mu$, while above the energy scale $\mu$, the term 
$\sigma \pont /({4\mu })$  would  ascend to certain renormalizable terms. 

In particular, the  term ${\sigma \pont }/({4\mu })$ can be 
generated by the gravitational axial anomaly \cite{Delbourgo:1972xb}. At energies above $\mu$, 
one starts with a gravitational theory of a massless fermion $\Psi$, 
coupled  to a complex scalar field $\Sigma$ with strength set by a Yukawa 
coupling $\lambda$, 
\be
\lambda\, ({\bar \Psi}_L \Sigma \Psi_R +{\bar \Psi}_R \Sigma^+ \Psi_L) \,.
\label{LR}
\ee
Furthermore, the complex scalar has its own conventional kinetic term 
and a quartic potential. All these term are symmetric with respect to  a global $U(1)$  
axial Peccei-Quinn (PQ) transformations
\be
\Sigma \to e^{i\beta } \Sigma, ~~~
\Psi_R \to e^{-i\beta/2 } \Psi_R, ~~~\, \Psi_L \to e^{i\beta/2 } \Psi_L\,.
\label{PQ}
\ee 
The PQ symmetry  is spontaneously broken by a  nonzero vacuum expectation value of the scalar
 $\langle \Sigma \rangle = \mu $, due to  the symmetry breaking scalar potential.  
Both the fermion and modulus of the scalar, $\rho = \sqrt {\Sigma^+\Sigma}$,  acquire their masses
due to the vacuum expectation value $\langle \Sigma \rangle = \mu $. These masses  
are proportional to the respective coupling 
constants and the energy scale $\mu$. Furthermore, these massive field  can be integrated out  
below their mass scales.  However, the phase of the scalar field, $\sigma$, remains massless  
as it is a Nambu-Goldstone (NG) mode of the spontaneously broken PQ symmetry.  
At low energies  only this massless state is kept, and   its low energy action 
can be deduced by substituting $\Sigma = \mu {\rm exp}{ (\sigma/\mu )}$ and  
calculating the anomalous diagrams, giving rise to the term proportional to ${\sigma \pont }/{4\mu }$.

We have already specified that  $\mu$ is considered to be much smaller than the Planck mass. 
The question though is how small can $\mu$ be. The fermion $\Psi$ and the scalar $\rho$ have masses 
proportional to $\mu$ and would  have been  accessible to accelerator experiments for $\mu < ~TeV$; 
however,  the fermion and  scalar do not couple to any other fields besides  gravity, and 
can only be produced in the accelerators via gravity mediated processes, which are very much suppressed
at energies below  $~TeV$.

For low values of $\mu$ one may worry about  nonlinear interactions of gravitons  
becoming strong at energies much lower that the Planck mass due to the 
new vertices introduced by Eq.~(\ref {CSsigmaCS}). For instance, the four graviton scattering 
amplitude of GR will be amended  by a set of new diagrams using the exchange of 
$\sigma$  due to the cubic vertex given by Eq.~(\ref {CSsigmaCS}). To  get a sense of the 
magnitude of  these corrections, we expand over a flat spacetime metric, $g_{ab}=\eta_{ab}+h_{ab}$, 
and  rescale $h\to h/M_P$,  to normalize canonically $h$'s kinetic term.  As a result,  we get 
from Eq.~(\ref {CSsigmaCS}) the scaling of the new cubic vertex
 \be
{\partial \sigma (\partial h) (\partial \partial h) \over \mu M_P^2}\,.
\label{cubic}
\ee     
Thus, the strong scale  is given by 
\be
\Lambda_s = (\mu M_P^2)^{1/3}\,.
\label{strong}
\ee 
Furthermore, the dCS term in Eq.~(\ref {CSsigmaCS}) will generate
higher order vertices, such as the one containing $\partial\sigma$ and three powers of
$\partial h$, but those terms will be suppressed by the scale, $ (\mu \, M_P^3)^{1/4}$, which is
higher than $\Lambda_s$.  All other higher vertices obtained  from Eq.~(\ref {CSsigmaCS})  will give 
even higher scales, and hence, $\Lambda_s$ is the lowest one to worry about.\footnote{One can also realize the Chern-Simons term from a BF theory perspective, see e.g. \cite{Alexander:2005vb}.}

For the value of $\mu$ as astonishingly small  as the present day
Hubble constant, $\mu \sim H_0\sim 10^{-33}\, eV$, the corresponding value of the strong scale 
is $\Lambda_s \sim 5\,\cdot 10^{7}\, eV$. The latter  is much higher than the scale of 
$10^{-2} eV$, up to  which precision gravity measurements have so far probed deviations 
from conventional gravity.  
%Hence, in the perturbative effective theory approximation, the range of scales we could  consider is $10^{-33}\, eV < \mu << M_P$.

 What is the range of $\mu$ in dCS that is allowed by non perturbative physics? The most stringent constraint on dCS gravity to date was established in Ref.~\cite{Silva:2020acr}, which used the gravitational wave data obtained by the LIGO/Virgo collaboration for the merger of two neutron stars~\cite{TheLIGOScientific:2017qsa}, as well as the X-ray data obtained by the NICER collaboration for the pulse profile emitted by a rotating neutron star~\cite{Riley:2019yda,Miller:2019cac}. This constraint requires that $\mu \gtrsim 4 \times 10^{-50} eV$ (or $\alpha^{1/2} \lesssim 8.5$km, in the notation of the next section) to 90\% confidence. Note in passing that binary pulsar observations cannot yet be used to place stringent constraints on dCS gravity because such binaries are widely separated, and thus, the Pontryagin source to the pseudo-scalar field is too small~\cite{Yagi:2013mbt}.

% This question will be discussed  in the context of rotating black holes in the  subsequent sections. Here we just make one comment. 
 
Before moving on, let us make a final comment about the mass of the pseudoscalar field $\sigma$.  There is no mass or potential terms associated with the pseudoscalar  field $\sigma$ 
 in Eq.~(\ref {CSsigma}), since   $\sigma$ is a NG boson of spontaneously broken PQ symmetry. 
 In the perturbative approximation, quantum corrections will not generate a nonzero mass and potential for $\sigma$ 
because of the shift symmetry, $\sigma\to\sigma +const$. However,  this symmetry is expected to be 
broken  by non-perturbative  quantum gravity effects and the pseudoscalar field would then 
acquire a mass \cite{Kallosh:1995hi}. The induced mass can be estimated; when the saddle-point  
approximation for the quantum gravity path integral is justified, the induced mass  ends up being 
small by an exponential factor, $e^{- x}$ with $x >> 100$ and $\mu<<M_P$ \cite{Kallosh:1995hi}. 
In what follows we will  consider distance  scales much shorter than the inverse of the induced 
pseudoscalar mass, and hence we will  ignore the small induced mass term in the action.

%%%%%%%%%%%%%%%%%%%%%%%%%%%%%%%%%%%%%%%%
\section{CS Caps for Rotating Black Holes}
\label{ABC}
 
Let us rescale  the  $\sigma $ field in Eq.~(\ref {CSsigma}) as $\sigma = M_P \, \vartheta$,
and pull out the overall factor of $M_P^2$ in front of the action; having done that, 
let us set   $M_P=1/ \sqrt{G_N}=1$.  In these geometric units the dCS gravity action reads~\cite{jackiw}:
\be
\label{CSaction}
S =  \int d^4x \sqrt{-g} \left[\kappa R + \frac{\alpha}{4} \vartheta \; \pont
- \frac{1}{2} \left(\nabla_a \vartheta\right) \left(\nabla^a \vartheta\right) \right]\,, 
\ee
where $\kappa = (16 \pi)^{-1}$,  and $\alpha \equiv 1/(\mu M_P)$. Therefore, although $\mu$ has units of energy in natural units, $\alpha$ has units of km$^2$ in geometric units.
%, which equals to $1/\mu$ in our units. 

The modified field equations can be obtained by varying the action in Eq.~(\ref{CSaction}) 
with respect to the metric: 
\begin{align}\label{eq:MetricEOM1}
G_{ab} + \frac{\alpha}{\kappa} \, C_{ab} &= \frac{1}{2 \kappa} T_{ab}\,,
\end{align}
where $G_{ab}$ is the Einstein tensor, and the traceless `C-tensor' is defined as
\begin{align}
C^{ab} = \left(\nabla_{c} \vartheta \right) \; \epsilon^{cde(a} \nabla_{e} R^{b)}{}_{d} + \left(\nabla_{c} \nabla_{d} \vartheta \right) \; {}^{\ast}R^{d(ab)c}\,.
\label{eq:C-tensor}
\end{align}
The stress-energy tensor for the scalar is 
\begin{align}
T_{ab} := \left(\nabla_{a} \vartheta \right) \left( \nabla_{b} \vartheta \right) - \frac{1}{2} g_{ab} \left(\nabla_{c} \vartheta \right) \left(\nabla^{c} \vartheta \right)\,,
\end{align}
and we will assume that apart from this scalar field the spacetime is empty. Variation of the action with respect to the scalar field yields its evolution equation
\begin{align}
\square \vartheta &= -\frac{\alpha}{4 \kappa} \, \pont\,,
\label{eq:theta-evolution}
\end{align}
where $\square$ stands for the d'Alembertian operator. 

The field equations are given by Eqs.~(\ref{eq:MetricEOM1}) and~\eqref{eq:theta-evolution}, but they simplify somewhat in trace-reversed form: 
\be
 R_{ab}  = 8 \pi \bar{T}_{ab} - 16 \pi \alpha C_{ab},
\label{eq:eom}
\ee
because the C-tensor is traceless, $C^a{}_a=0$, where
\be
\bar{T}_{ab} := \left(\nabla_{a} \vartheta \right) \left( \nabla_{b} \vartheta \right)\,,
\ee
is the trace-reversed stress-energy tensor of the scalar field. From this formulation, it is clear that in the pure vacuum case, i.e.~when $\bar{T}_{ab}=0$, then the pseudo-scalar field must be a constant and dCS gravity reduces continuously to GR. 

When the so-called Pontryagin constraint holds on a subspace of solutions, i.e.~when $\pont=0$ on shell, then dCS gravity simplifies significantly. One can show that $\pont=0$ for any spherically symmetric spacetime, regardless of whether it is static or not~\cite{Grumiller:2007rv, Shiromizu:2013pna}. If so, the pseudo-scalar field then satisfies an unsourced wave equation, $\square \vartheta = 0$. If one imposes a ``no-cosmological scalar field'' boundary condition, i.e.~$\nabla_a \vartheta = 0$ at spatial infinity, then $T_{ab} = 0 = C_{ab}$ for stationary solutions. In this case, all spherically symmetric, stationary, spacetimes must be Ricci flat, and one concludes that all spherically symmetric, vacuum solutions in dCS gravity must be identical to those in GR~\cite{Grumiller:2007rv}. In particular, this implies that the Schwarzschild metric continues to be a solution of dCS gravity. 

When we consider spacetimes that break spherical symmetry, however, then the Pontryagin density does not vanish and GR solutions will not be solutions of dCS gravity. For example, when considering spacetimes that are stationary but axisymmetric, then the Pontryagin density sources a non-trivial scalar field, which then back-reacts on the metric to induce non-GR modifications. Such an analysis can be carried out to find slowly-rotating black hole solutions in dCS gravity, as done first in~\cite{Yunes:2009hc}, and then extended to second-order and fifth order in rotation in~\cite{Yagi:2012ya} and~\cite{Maselli:2017kic} respectively. 

Let us consider the dCS gravity solution that represents a stationary and axisymmetric spacetime~\cite{Yunes:2009hc,Yagi:2012ya,Maselli:2017kic} valid to fifth order in a slow-rotation expansion in Boyer-Lindquist-like coordinates, with ADM angular momentum $J_{\rm ADM} = M a$ and ADM mass $M_{\rm ADM} = M$. To leading order in $a/M \ll 1$, the modified rotating black-hole solution in dCS gravity using Boyer-Lindquist-like coordinates $(t,r,\theta,\phi)$ is
\begin{align}
ds^{2} \!&= \!ds^{2}_{K} \!+\! \frac{5}{4}\zeta M \chi \frac{M^4}{r^4} \left( 1+\frac{12}{7}\frac{M}{r} + \frac{27}{10} \frac{M^2}{r^2} \right) \sin^2\!\theta dt d\phi\,,
\label{metric-linear}
\end{align}
where $ds^{2}_{K}$ is the Kerr solution of general relativity, $\chi = a/M$ is the dimensionless spin parameter, and
\be
\zeta = \frac {16 \pi \alpha^2}{(G_N M)^4}\,,
\label{zeta}
\ee
is a dimensionless coupling parameter.
The higher order in spin terms introduce modifications to all other components of the metric that can be found in~\cite{Maselli:2017kic}, but their expressions to ${\cal{O}}(a^5/M^5)$ are long and un-illuminating, so we will not present them here.

Let us us consider the behavior of both null and timelike geodesics in this spacetime. First, consider a static timelike observer in this spacetime. The tangent to such an observer's geodesic is $k^a_{\static} = \gamma_{\static} [1,0,0,0]$ and $\gamma_{\static}$ is a normalization constant to ensure $k_a^{\static} k^a_{\static} = -1$. With that in hand, employing Maple and the GRTensorIII software \cite{grtensor}, we calculate the quantity whose sign defines whether the 
geodesic congruences are converging or diverging: 
\begin{align}
R_{a b} \; k_{\static}^a k^b_{\static} &= \frac{45}{4} \; \zeta \chi^2 \; f  \frac{\gamma^8}{M^{2}} 
\left[
1 + 2 c_{\theta}^2 + \frac{40\gamma}{15} \left(1 + \frac{3}{4} c_{\theta}^2 \right)
\right. 
\nonumber \\
&\left.
+ 6 \gamma^2 \left(1 + \frac{1}{3}  c_{\theta}^2\right) - \frac{312}{5} \gamma^3 c_{\theta}^2
\right] + {\cal{O}}(\zeta \chi^4)\,,
\label{Rkk}
\end{align}
where $f := 1 - 2 \gamma$, $\gamma := M/r$, and $c_\theta := \cos{\theta}$. 
Similarly, consider a null observer with tangent vector $l^a = [l, g(r,\theta), 0,0]$, where $g(r,\theta)$ is a function such that the null condition $l_al^a=0$ is satisfied. In this case we have:
\begin{align}
    R_{a b} \; l^a l^b &= \frac{25}{32} \; \zeta\chi^2 \; f \frac{\gamma^6}{M^2} \left[c_\theta^2 + 4\gamma c_\theta^2 
  + \frac{72}{5}\gamma^2 \left(1 + \frac{43}{24}c_\theta^2\right) \right. \nonumber \\ &\left. + \frac{192}{5}\gamma^3 \left(1 + \frac{13}{32}c_\theta^2\right) 
  + \frac{432}{5}\gamma^4\left(1 - \frac{1}{15}c_\theta^2\right)\right. \nonumber \\ &\left.  - \frac{19872}{25}\gamma^5 c_\theta^2\right] +{\cal{O}}(\zeta \chi^4)\,,
  \label{Rll}
\end{align}
One can check by direct evaluation that both of these quantities are positive definite almost everywhere. In those regions 
the geodesics will be focusing. However, there are  regions where the quantities in Eq.~(\ref {Rkk}) and Eq.~(\ref {Rll}) are negative. 
 In particular, when one looks at spacetime regions near the polar axis (i.e.~near $\theta = 0$) and close to the horizon, one finds that the above contractions changes their signs. Note that, as we will show below, the location of the ergosphere coincides with the location of the event horizon along the polar axis in this solution, just like it does for the Kerr metric, so static observers do exist right outside the horizon along the polar axis. 

Let us discuss these unusual regions further. The considered solution is known to have an event horizon located at
\be
r_{\EH} = r_{\EH,\K}  - \zeta M \left[\frac{915}{28672} \chi^2 + \frac{351479}{13762560} \chi^4 + {\cal{O}}(\zeta \chi^6)\right]\,,
\ee
and an ergosphere whose outer edge is located at
\begin{align}
r_{\ergo} &= r_{\ergo,\K}  - \zeta M \left[
\left( 
\frac{915}{28672} + \frac{709}{7168} s_{\theta}^2 
\right) \chi^2 +
\right.
\nonumber \\
&\left.
\left( 
 \frac{351479}{13762560} 
- \frac{336421}{2408448} s_{\theta}^2
+ \frac{151229}{1605632} s_{\theta}^4
\right) \chi^4 
\right.
\nonumber \\
&\left.
+ {\cal{O}}(\zeta \chi^6)\right]\,,
\end{align}
where $s_{\theta} := \sin{\theta}$, while $r_{\EH,\K} = M + M (1 - \chi^2)^{1/2}$ and $r_{\ergo,\K} = M + M (1 - \chi^2 c_{\theta}^2)^{1/2}$ are the locations of the event horizon and the outer-edge of the ergosphere for the Kerr metric respectively \cite{Yunes:2009hc,Yagi:2012ya,Maselli:2017kic}; notice that the outer edges of the ergosphere coincide on the polar axis $\theta = 0$. 

\begin{figure*}[htb]
\includegraphics[width=0.46\linewidth]{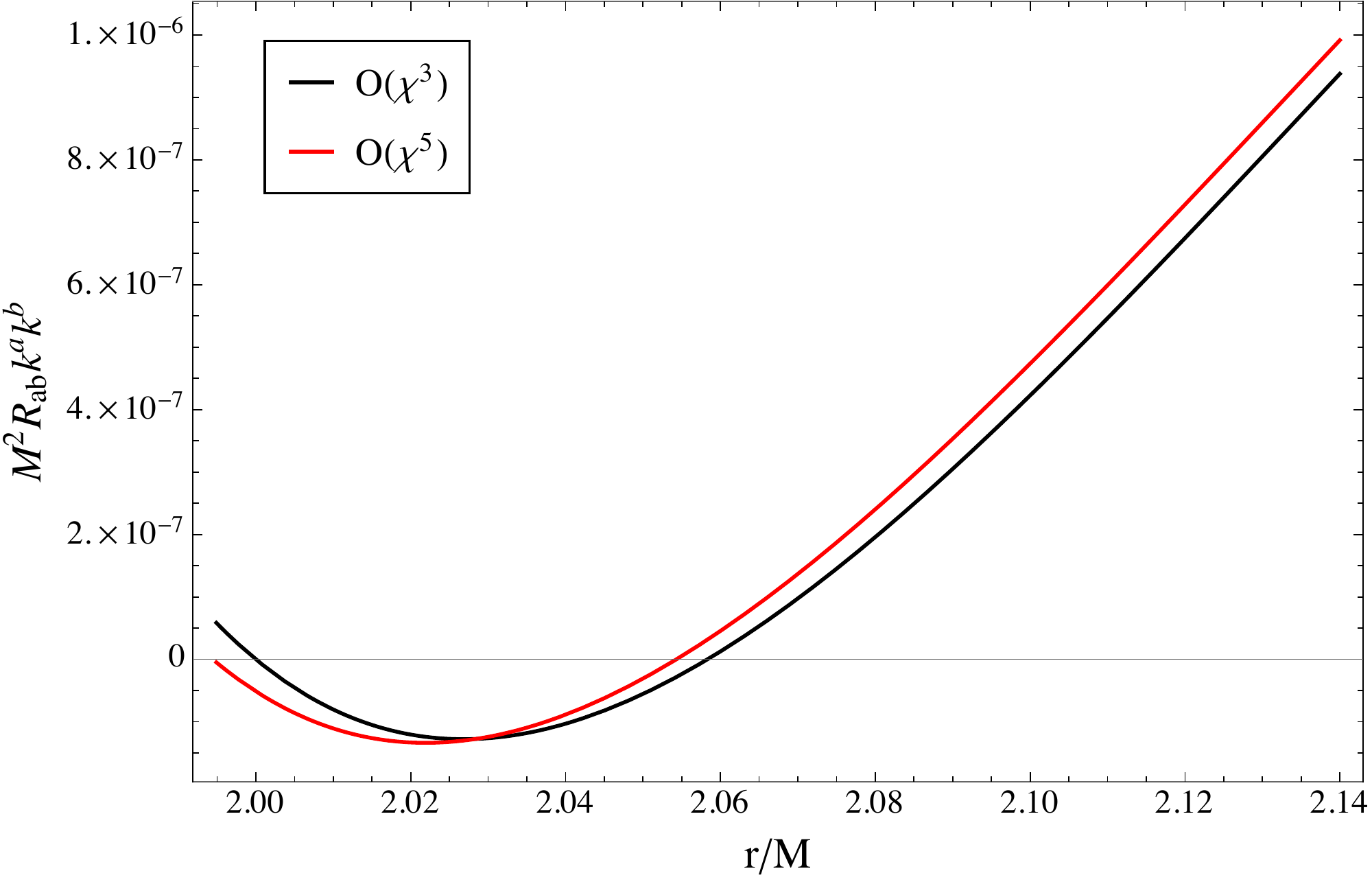}
 \includegraphics[width=0.52\linewidth]{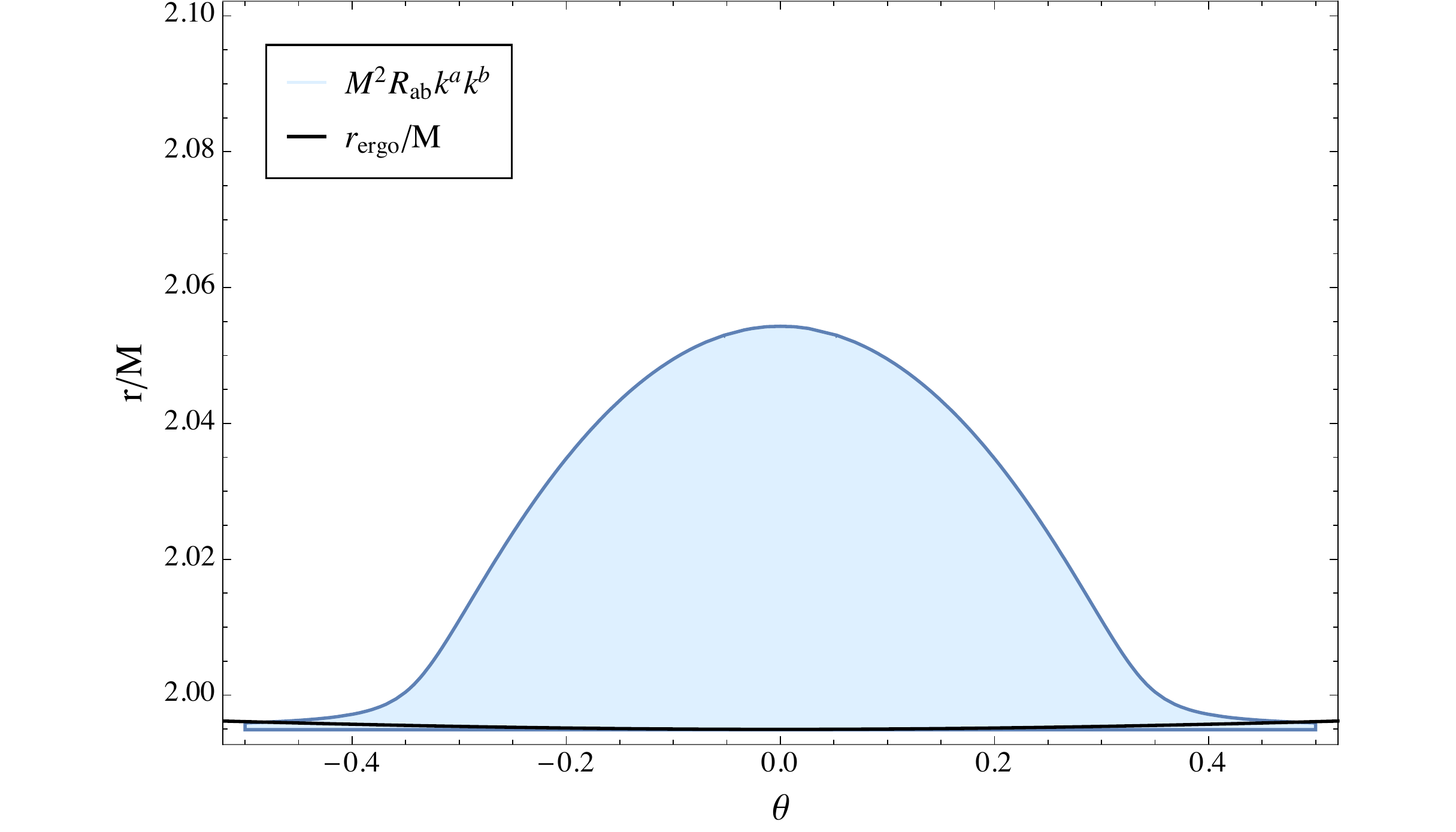}\\
 \includegraphics[width=0.46\linewidth]{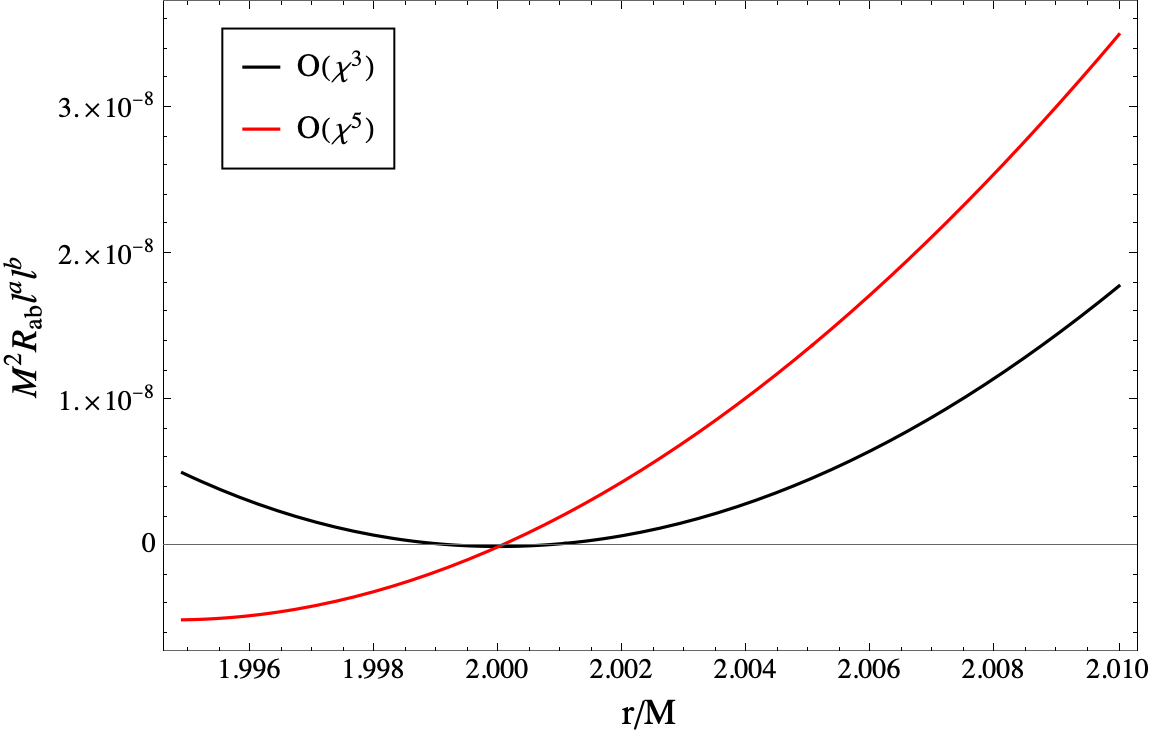}
  \includegraphics[width=0.52\linewidth]{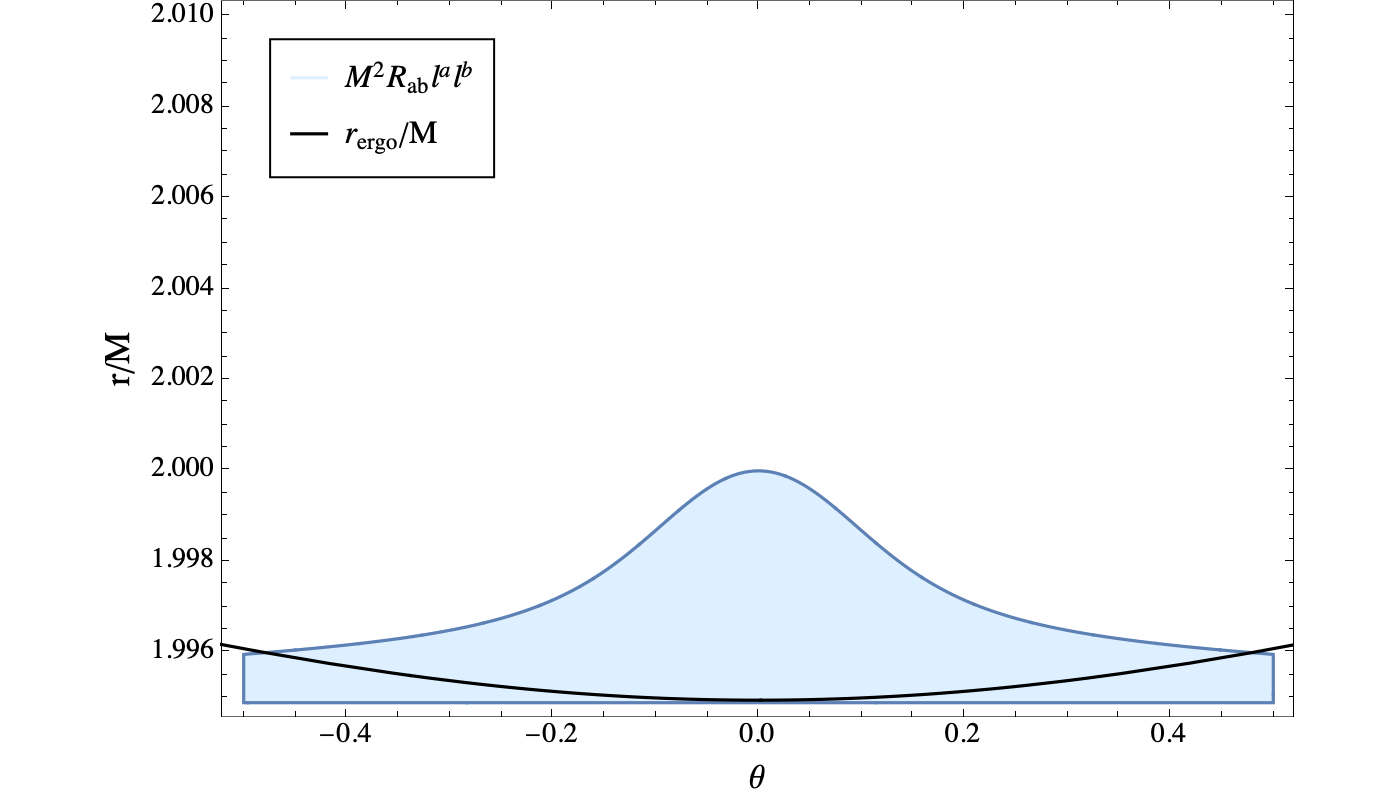}
\caption{(color online) Dimensionless contraction of the Ricci tensor with the tangent vector of  timelike (top) and null (bottom) congruences $M^2 E[k] = M^2 R_{ab} k^a k^b$ and $M^2 E[k] = M^2 R_{ab} l^a l^b$ computed along the ($\theta=0$) polar axis (left panels) and in the $r$--$\theta$ plane (right panels). In the left panels, the red and blue curves denote $M^2 E[k]$ calculated with a BH solution in dCS gravity to third- and fifth-order in rotation, respectively, both for a black hole with dimensionless spin $\chi = 0.1$ and dimensionless dCS coupling $\zeta = 0.1$. In the right panels, we present $M^2 E[k]$ computed with a fifth-order in rotation dCS BH metric (blue shaded region) assuming $\chi = 0.1$ and $\zeta = 0.1$, and for comparison, we also present the outer edge of the ergosphere in dCS gravity. Observe that the contractions $R_{ab} k^a k^b$ and $R_{ab}l^al^b$ switch sign close to the horizon, in a cup-shaped region around the polar axis. Note that the x-axis on the right panels is given in $\theta$, not in $r/M$.
}\label{fig:Rabkakb}
\end{figure*}
The top left panel of Fig.~\ref{fig:Rabkakb} shows the contraction $R_{ab} k^a k^b$ (timelike geodesics) to third- and to fifth-order in the slow-rotation approximation on the polar axis close to but outside of the event horizon. Observe that this contraction flips sign, regardless of the order of the  slow-rotation approximation used for the metric. Extending this analysis to all $\theta$, we see that indeed there is a cone shaped region near the horizon extending from the pole up to $\sim 20^\circ$ on either side where the sign of the contraction $R_{ab}k^ak^b$ flips and becomes negative. This is shown in the top right panel of Fig.~\ref{fig:Rabkakb} (blue region) with a metric valid to to $\mathcal{O}(\chi^5)$, where we also include the location of the outer edge of the ergosphere (black curve) for comparison. 

The bottom left and right panels show the contraction $R_{ab} l^al^b$ (null geodesics) on the polar axis to third- and fifth-order in rotation, and extended to all $\theta$, respectively. We see that there is a similar, much smaller region close to the horizon where the null contraction flips sign. Note that in order to obtain the negative region, one must go to $\mathcal{O}(\zeta\chi^5)$ in perturbation. We expect that going to higher orders in the perturbative expansion will yield small corrections and will not change the negative result. %In this case, if one considers radially infalling geodesics, we find that the geodesics slow down very slightly in the dCS case as compared to Kerr, but overall follow the same trajectories. } 

The size of the regions of interest do not depend on the value of $\zeta$, which is proportional to the coupling $\alpha^2$, see Eq.~(\ref{zeta}). In order to remain in the regime of validity of the slow rotation approximation, we must have $\zeta \ll 1$, so we have taken $\zeta = 0.1$ as a representative example in Fig.~\ref{fig:Rabkakb}. Varying this value changes the magnitude of negativity for the contractions $R_{ab}k^ak^b$ and $R_{ab}l^al^b$ in the CS caps, but it does \textit{not} change the boundaries of the caps. That is, we have explored versions of Fig.~\ref{fig:Rabkakb} evaluated with various values of $\zeta$ (e.g.~$\zeta = (10^{-3},10^{-2},10^{-1})$) and in all cases the change in sign occurs roughly along the boundary of the same caps shown in the right panel of that figure (although how negative the contraction is does scale with $\zeta$).

This technical point -- that we are necessarily constrained to consider small values of $\zeta$ because 
of the lack of an exact solution --  also forces us to consider black holes that are not too small %constrains the masses of the  Black Holes for  which our calculations can reliably predict the existence of the sizable Chern-Simons caps. 
To see this, let us use the definition in Eq.~(\ref{zeta}), and require that $\zeta \leq 0.1$ to find %with the assumed value, $\zeta=0.1$;
\be
% M \geq \left(\frac{16 \pi}{\zeta}\right)^{1/4} \sqrt{\alpha} \approx 3 M_\odot \left(\frac{0.1}{\zeta}\right)^{1/4} \left(\frac{\alpha}{1 \; {\rm{km}}}\right)^{1/2}
M \geq 3 M_\odot  \left(\frac{\alpha}{1 \; {\rm{km}}}\right)^{1/2} , 
\ee
or restoring the  powers of the Planck mass
\begin{align}
% M &\geq \left(\frac{16 \pi}{\zeta}\right)^{1/4} \!\!\! M_p \sqrt{\frac{M_p}{\mu}}  \approx 2 M_{\odot} \left(\frac{10^{-47} \; eV}{\mu}\right)^{1/2}\,.
M &\geq  2 M_{\odot} \left(\frac{10^{-47} \; eV}{\mu}\right)^{1/2}\,.
%5\,M_p \,\sqrt{{M_p\over \mu}}\,.
\label{M-zeta}
\end{align}
If Nature is described by dCS gravity as an effective theory, then depending on Nature's value of $\mu$ (or $\alpha$), our calculations would be valid for black holes of a different mass. More specifically, for our calculations to be valid, the smaller Nature's $\mu$ is (or the larger Nature's $\alpha$ is), the heavier the black holes we can consider would have to be.
%and predict the appreciable size CS caps.   {\ny{For instance, 
%saturating the constraint on $\mu$ from LIGO/Virgo and NICER observations, $\mu \gtrsim 4 \times 10^{-50}\,eV$,  we estimate, $M\gtrsim 3 \times 10^{58} \,GeV \approx 28 M_\odot$. 
%which is of the order of the mass  of the planet Mercury. 
%More stringent astrophysical constraints that bound $\mu$ to higher values (or $\alpha$ to smaller values) would lead to smaller values of $M$.}}

The above limitation on the BH mass is  a result of our small coupling approximation (ie.~$\zeta \ll 1$) when finding slowly rotating 
BH solutions.  However, it might well be  that the CS caps are universal  for all rotating black holes 
in the dCS theory. Such an outcome is not ruled out  by  general arguments  of continuity in the value of the 
parameter $\zeta$,  and by the considerations of the focusing theorem in the next section showing that the 
dCS can in general permit negative values for the contraction in Eq.~(\ref {Rkk}), irrespective of 
the approximation used.

% \begin{figure*}
%     \centering
%     \includegraphics[width=\linewidth]{RkkO3}
%         \includegraphics[width=\linewidth]{RkkO5}
%     \caption{Region of negative $R_{ab}k^ak^b$ up to third order in rotation.}
%     \label{fig:RkkO3}
% \end{figure}
% \begin{figure}
%     \centering
%     \includegraphics[width=\linewidth]{RkkO5}
%     \caption{Region of negative $R_{ab}k^ak^b$ up to fifth order in rotation.}
%     \label{fig:RkkO4}
% \end{figure*}

\begin{figure}
\includegraphics[clip=true,width=1.1\linewidth]{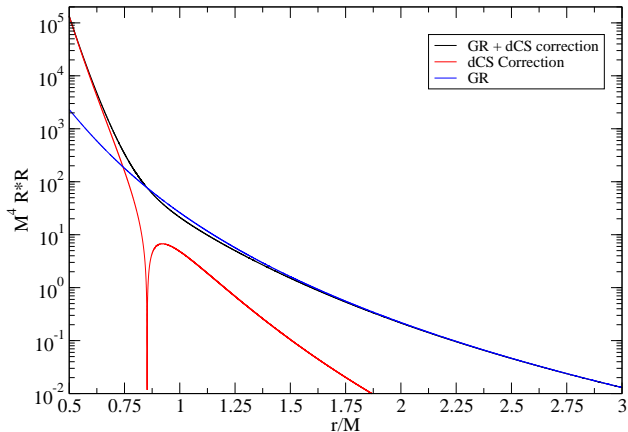}
\caption{(color online) Dimensionless Pontryagin density of a slowly-rotating black hole solution computed to fifth order in slow rotation in GR (blue), in dCS gravity (black) and using only the $\alpha^2$ correction to the GR solution in dCS gravity (red) for a black hole with spin $\chi = 0.1$ and dCS coupling $\zeta = 0.1$. Observe that the $\alpha^2$ correction to the Pontryagin density is much smaller than the GR contribution outside the horizon ($r/M \gtrsim 2$), which is precisely where $R_{ab} k^a k^b$ becomes negative.}\label{fig:Pont}
\end{figure}

One may be worried that the curvatures close to the horizon are so large that we are exploring these slowly-rotating solutions outside of the regime of validity of the effective field theory, where these solutions are calculated in the first place. The cut-off scale of the theory, i.e.~the scale inside which the small-GR-deformation approximation of effective field theory breaks down, can be approximated by computing the Pontryagin density with the approximate black hole solutions. Figure~\ref{fig:Pont} presents the Pontryagin density computed with the Kerr metric  and with the dCS metric for a slowly-rotating black hole. Observe that the dCS correction to the Pontryagin density, i.e.~the term proportional to $\alpha^2$ in the calculation of $R^{*}R$ with the dCS metric,  exceeds the GR value only deep inside the event horizon for $r/M < 0.75$, and nowhere near the regime where the contraction, $R_{ab} k^a k^b$ flips sign (which from Fig.~\ref{fig:Rabkakb} we recall occurs for $2.06 \lesssim r/M \gtrsim 2$ for timelike geodesics).\footnote{Even though we are working here in Boyer-Lindquist coordinates, the statement that the curvature becomes large only deep inside the horizon is also true in horizon-penetrating coordinate systems.}   

%One may also worry that the above results hold only for static observers, and may actually not hold for other observers. For a stationary observer with $k^a_{\stationary} = \gamma_{\stationary} [1,0,0,\Omega]$, $\gamma$ another normalization constant to ensure $k_a^{\stationary} k^a_{\stationary} = -1$, and $\Omega$ the angular velocity of the observer, we find that
%
%\begin{align}
%R_{a b} \; k^a_{\stationary} k^b_{\stationary} &= 
%\left[
%1 + \right] + {\cal{O}}(\zeta \chi^4)\,,
%\end{align}
%
%Finally, for a generic observer on a timelike geodesic with $k^a_{\generic} = [\dot{t},\dot{r},\dot{\theta},\dot{\phi}]$, where the overhead dots stand for partial derivatives with respect to proper time, we find that
%
%\begin{align}
%R_{a b} \; k^a_{\generic} k^b_{\generic} &= 
%\left[
%1 + 
%\right] + {\cal{O}}(\zeta \chi^4)\,,
%\end{align}
%where we have used the geodesic equations.
%

An agreement emerges between our result and both analytic~\cite{Yunes:2009hc,Yagi:2012ya,Maselli:2017kic} and numerical~\cite{Delsate:2018ome,Sullivan:2020zpf} studies that have found black hole spacetimes. Those studies found stationary and axisymmetric solutions in dCS gravity that differ from the Kerr spacetime everywhere on the manifold, leading for example to dCS corrections to the location of the event horizon and the ergosphere. These solutions agree with the Kerr metric in that they both posses spacetime regions inside their respective event horizons where curvature invariants diverge, which is why we say these solutions represent black holes. The dCS corrections to the metric, however, become dominant over the GR terms inside the event horizon, as shown in Fig.~\ref{fig:Pont}, and in particular before reaching the singularity. As usual, the curvature singularities in these modified solutions are outside the validity of the effective theory. It is entirely possible that these singularities could then be cured by higher order terms in the action that have been neglected. It would be interesting to see if higher curvature corrections related to a stringy excitations in higher dimensions  may become relevant near the singularity to resolve it, in the spirit of topological stars \cite{Bah}.

%%%%%%%%%%%%%%%%%%%%%%%%%%%%%%%%%%%%%%%%%%%%%%%%%%%%%%%%%%%%%%%%%%%%%%
\section{Geodesic (de)focusing and dCS Gravity}

%---------------------------------------------------------------
\subsection{Focusing and the Hawking-Penrose Theorem}

To review the well-known focusing theorem~\cite{1955PhRv...98.1123R},  one starts from the Ricci identity,
\be 
(\nabla_{a}\nabla_{b} - \nabla_{b}\nabla_{a})k^{a} = R^{a}{}_{cab} k^{c},
\ee
and derives the Raychaudhuri equation~\cite{1955PhRv...98.1123R},
\be 
\dot{\Theta} = -\sigma_{ab}\sigma^{ab}- \frac{1}{3} \Theta^{2} - E[k]\,,
\ee
for vorticity-free congruences of non-intersecting world lines. The quantity $\sigma_{ab}$ is the shear tensor of this congruence, while $\Theta=\nabla_{c}k^{c}$ is the expansion scalar, and $E[k] := R_{cd}k^{c}k^{d}$, where $k^c$ is the timelike or null tangent vector field of the congruence. 

The focusing theorem for timelike geodesics states that if the strong-energy condition $\bar{T}_{ab} k^a k^b > 0$ holds, then vorticity-free geodesics will focus in GR. This result is established directly from the  Raychaudhuri equation presented above, upon the use of the Einstein equations to write $E[k]= 8 \pi \bar{T}_{cd}k^{c}k^{d}$, where $\bar{T}_{ab}$ is the trace-reversed stress-energy tensor. Using the Einstein equations and the strong-energy condition, it is obvious that the right-hand side of the Raychaudhuri equation is negative because $E[k] > 0$, which then means that the rate of change of the expansion scalar with respect to the geodesic's affine parameter is negative, and the worldlines will focus within a finite value of the affine parameter, reflecting the attractive nature of gravity. An analogous argument holds for null geodesics, provided that the null energy condition is satisfied.

The focusing theorem leads to the Hawking-Penrose singularity theorem as follows. Consider two events $A$ and $B$ in a globally hyperbolic spacetime which contains a trapped surface, that are connected via a timelike or null curve. If this is the case, there must exist a geodesic of maximal length $\gamma$ that connects these two points, on which there are no conjugate points. The focusing theorem, however, establishes that all geodesics emanating from $A$ will focus in finite affine parameter, leading to conjugate points. Intuitively, a geodesic cannot be extended beyond a conjugate point, and therefore one cannot reach point $B$, so the spacetime must be geodesically incomplete. In summary~\cite{Hawking:1969sw}:

\bf{Theorem 1}\rm(Hawking-Penrose singularity theorem).\it{If a globally-hyperbolic spacetime contains a non-compact
Cauchy hypersurface $\Sigma$ and a closed future-trapped surface, and if the convergence condition, $R_{ab}u^{a}u^{b} \geq 0$  holds for null $u^{a}$, then there are future incomplete null geodesics.}\rm

Einstein's theory of general relativity predicts that singularities are unavoidable, since the energy momentum tensor along timelike or null geodesics, will be positive definite according to the Einstein field equations if the strong energy condition holds. On the other hand, consistent modifications of general relativity that violate the strong energy condition can lead to violations of the focusing theorem, and therefore the evasion of singularities. Clearly, if the discriminant $E[k] < 0$, then it could be that $\dot{\Theta} >0$, which would imply that geodesics defocus, leading to gravitational repulsion and the possibility of an evasion of singularities. We have shown in Section ~\ref{ABC} that the slowly rotating dCS solution contains geodesics for which $E[k] <0$ close to the black hole.Note, however that due to the smallness of the dCS caps, both the null and strong \textit{average} energy conditions \cite{Borde_1987} remain satisfied. In what follows, we show that dCS gravity naturally has the mathematical and physical features needed to violate the conditions of the focusing theorem. 

%---------------------------------------------------------------
\subsection{The Focusing  and Hawking-Penrose Theorems in dCS Gravity}

Recall that both the focusing and Hawking-Penrose theorem rely on satisfying the constraint $R_{ab}k^{a}k^{b} \geq 0$ with $R_{ab}$ the Ricci tensor and $k^a$ the tangent to a timelike or a null geodesic congruence. Using the modified field equations in dCS gravity, this condition becomes,
\be 
\bar{T}_{ab} k^{a}k^{b} \geq 2 \alpha C_{ab} k^{a}k^{b}\,, 
\ee 
and we see that in general this need not be satisfied allowing for the possibility that timelike or null congruences will defocus, avoiding a singularity.  

When precisely do we have a violation of the focusing theorem? Using the definition of the scalar-field stress-energy tensor and the C-tensor, this occurs when
\begin{align}
\label{ineq}
& 2 \alpha \left(\nabla_{c} \vartheta \right) \; \epsilon^{cdea} \left(\nabla_{e} R^{b}{}_{d}\right) k_{a} k_{b}
\nonumber \\
&+ 2 \alpha \left(\nabla_{c} \nabla_{d} \vartheta \right) \; {}^{\ast}R^{dabc} k_{a}k_{b} > \left(\nabla_{a} \vartheta \right) \left( \nabla_{b} \vartheta \right) k^{a}k^{b}\,,
\end{align}
where we have removed the symmetrization parenthesis in the C-tensor because we are contracting it with the symmetric tensor $k_a k_b$. We recognize the right-hand side as a total square of the directional derivative of the scalar field in the direction of the tangent to the congruence. Therefore, a sufficient (but not necessary) conditions for the Hawking-Penrose theorems to be violated is simply
\begin{align}
\label{ineq2}
&\left(\nabla_{c} \nabla_{d} \vartheta \right) \; {}^{\ast}R^{dabc} k_{a}k_{b} > \left(\nabla_{c} \vartheta \right) \; \epsilon^{dcea} \left(\nabla_{e} R^{b}{}_{d}\right) k_{a} k_{b}\,,
\end{align}
where we have eliminated the minus sign through a permutation of the indices in the Levi-Civita tensor. 

The above inequality is the best one can do without using additional approximations. Note that the equation of motion for the scalar field cannot be used to simplify the left-hand side of the above equation, because the double covariant derivative acting on the scalar field is not contracted into a d'Alembertian operator. A simplification one can do, however, is to work in an effective field theory approach. In the latter, one can substitute the modified field equations into the right-hand side of the above equations to find  
\begin{align}
\left(\nabla_{c} \nabla_{d} \vartheta \right) \; {}^{\ast}R^{dabc} k_{a}k_{b} &> \left(\nabla_{c} \vartheta \right) \; k_{a} k_{b} \; \epsilon^{dcea} 
\nonumber \\
& \times
\nabla_{e} \left(8 \pi \bar{T}^b{}_d - 16 \pi \alpha C^b{}_d\right) \,,
\end{align}
The right-hand side of this equation is quadratic in $\alpha$ because because the equation of motion of the scalar field $\vartheta$ is linear in $\alpha$. Therefore, to leading order in $\alpha$, the condition to ensure the Hawking-Penrose theorem is violated reduces to  
\begin{align}
&\left(\nabla_{c} \nabla_{d} \vartheta \right) \; {}^{\ast}R^{dabc} k_{a}k_{b} > 0\,.
\end{align}
There is no reason to expect that the left-hand side of the above equation will have a definite sign. Thus, in general, it would seem the assumptions required in the Hawking-Penrose theorem do not hold. Notice, moreover, that in the above derivation, we never used the fact that $k^a$ is the tangent vector to a timelike geodesic congruence, and so, this result also applies to null geodesic congruences. We then conclude that the usual proof of the Hawking-Penrose theorem does not go through in dCS gravity.
%using its and thus, it may be that in certain cases the theorem does not go through in dCS gravity. 

The lesson we have learned from this first study is that timelike and null geodesics can, and in general will defocus in dCS gravity, yielding a violation of the conditions required by the Hawking-Penrose singularity theorem. By itself, this does \textit{not} however mean that dCS gravity resolves the black hole singularity. At shorter distances, the dCS term would transform into the terms that gave rise to it in the low-energy approximation, as was was discussed in Section 2.  To fully understand this, we plan to study Oppenheimer-Snyder-like collapse of a rotating dust ellipsoid and neutron star collapse in dCS gravity and its short-distance completion to see if, when and how a singularity forms~\cite{AJY}. Moreover, in another upcoming study we will analyze solutions, including dynamical black hole and cosmological solutions, that obey the $R_{ab} k^a k^b < 0$ condition, and explore the prospect of what new physical effects this violation is signaling, including a potential new regime of super-radiance outside the rotating black holes \cite{AJGY}.  

\section{Acknowledgment}
We especially thank Jim Simons for proposing to look at the rotating Black Holes and  Hawking-Penrose theorem in the context of dynamical Chern-Simons gravity. We thank Will Farr and David Spergel for their useful comments and for looking at a draft of this work. GG is supported by NSF grant PHY-1915219.
%%%%%%%%%%%%%%%%%%%%%%%%%%%%%%%%%%%%%%%%%%%%%%%%%%%%%%%%%%%%%%%%%%%%%
\normalem
\bibliographystyle{apsrev}
\bibliography{master}

\begin{thebibliography}{33}
\expandafter\ifx\csname natexlab\endcsname\relax\def\natexlab#1{#1}\fi
\expandafter\ifx\csname bibnamefont\endcsname\relax
  \def\bibnamefont#1{#1}\fi
\expandafter\ifx\csname bibfnamefont\endcsname\relax
  \def\bibfnamefont#1{#1}\fi
\expandafter\ifx\csname citenamefont\endcsname\relax
  \def\citenamefont#1{#1}\fi
\expandafter\ifx\csname url\endcsname\relax
  \def\url#1{\texttt{#1}}\fi
\expandafter\ifx\csname urlprefix\endcsname\relax\def\urlprefix{URL }\fi
\providecommand{\bibinfo}[2]{#2}
\providecommand{\eprint}[2][]{\url{#2}}

\bibitem[{\citenamefont{Jackiw and Pi}(2003{\natexlab{a}})}]{Jackiw:2003pm}
\bibinfo{author}{\bibfnamefont{R.}~\bibnamefont{Jackiw}} \bibnamefont{and}
  \bibinfo{author}{\bibfnamefont{S.}~\bibnamefont{Pi}}, \bibinfo{journal}{Phys.
  Rev. D} \textbf{\bibinfo{volume}{68}}, \bibinfo{pages}{104012}
  (\bibinfo{year}{2003}{\natexlab{a}}), \eprint{gr-qc/0308071}.

\bibitem[{\citenamefont{{Alexander} and {Yunes}}(2009)}]{Alexander:2009tp}
\bibinfo{author}{\bibfnamefont{S.}~\bibnamefont{{Alexander}}} \bibnamefont{and}
  \bibinfo{author}{\bibfnamefont{N.}~\bibnamefont{{Yunes}}},
  \bibinfo{journal}{{Phys. Rep.}} \textbf{\bibinfo{volume}{480}},
  \bibinfo{pages}{1} (\bibinfo{year}{2009}), \eprint{0907.2562}.

\bibitem[{\citenamefont{Alvarez-Gaume and Witten}(1984)}]{ALVAREZGAUME1984269}
\bibinfo{author}{\bibfnamefont{L.}~\bibnamefont{Alvarez-Gaume}}
  \bibnamefont{and} \bibinfo{author}{\bibfnamefont{E.}~\bibnamefont{Witten}},
  \bibinfo{journal}{Nuclear Physics B} \textbf{\bibinfo{volume}{234}},
  \bibinfo{pages}{269} (\bibinfo{year}{1984}), ISSN \bibinfo{issn}{0550-3213}.

\bibitem[{\citenamefont{Polchinski}(2007)}]{Polchinski:1998rr}
\bibinfo{author}{\bibfnamefont{J.}~\bibnamefont{Polchinski}},
  \emph{\bibinfo{title}{{String theory. Vol. 2: Superstring theory and
  beyond}}}, Cambridge Monographs on Mathematical Physics
  (\bibinfo{publisher}{Cambridge University Press}, \bibinfo{year}{2007}), ISBN
  \bibinfo{isbn}{978-0-511-25228-0, 978-0-521-63304-8, 978-0-521-67228-3}.

\bibitem[{\citenamefont{Alexander et~al.}(2006)\citenamefont{Alexander, Peskin,
  and Sheikh-Jabbari}}]{PhysRevLett.96.081301}
\bibinfo{author}{\bibfnamefont{S.~H.~S.} \bibnamefont{Alexander}},
  \bibinfo{author}{\bibfnamefont{M.~E.} \bibnamefont{Peskin}},
  \bibnamefont{and} \bibinfo{author}{\bibfnamefont{M.~M.}
  \bibnamefont{Sheikh-Jabbari}}, \bibinfo{journal}{Phys. Rev. Lett.}
  \textbf{\bibinfo{volume}{96}}, \bibinfo{pages}{081301}
  (\bibinfo{year}{2006}),
  \urlprefix\url{https://link.aps.org/doi/10.1103/PhysRevLett.96.081301}.

\bibitem[{\citenamefont{Alexander and Gates}(2006)}]{Alexander:2004xd}
\bibinfo{author}{\bibfnamefont{S.~H.~S.} \bibnamefont{Alexander}}
  \bibnamefont{and} \bibinfo{author}{\bibfnamefont{J.}~\bibnamefont{Gates},
  \bibfnamefont{S.~James}}, \bibinfo{journal}{JCAP}
  \textbf{\bibinfo{volume}{0606}}, \bibinfo{pages}{018} (\bibinfo{year}{2006}),
  \eprint{hep-th/0409014}.

\bibitem[{\citenamefont{Green et~al.}(1987)\citenamefont{Green, Schwarz, and
  Witten}}]{Green:1987mn}
\bibinfo{author}{\bibfnamefont{M.~B.} \bibnamefont{Green}},
  \bibinfo{author}{\bibfnamefont{J.~H.} \bibnamefont{Schwarz}},
  \bibnamefont{and} \bibinfo{author}{\bibfnamefont{E.}~\bibnamefont{Witten}},
  \emph{\bibinfo{title}{Superstring Theory. Vol. 2: Loop Amplitides, Anomalies
  and Phenomenology}} (\bibinfo{publisher}{Cambridge University Press},
  \bibinfo{address}{Cambridge, UK}, \bibinfo{year}{1987}).

\bibitem[{\citenamefont{Jackiw and Pi}(2003{\natexlab{b}})}]{jackiw}
\bibinfo{author}{\bibfnamefont{R.}~\bibnamefont{Jackiw}} \bibnamefont{and}
  \bibinfo{author}{\bibfnamefont{S.~Y.} \bibnamefont{Pi}},
  \bibinfo{journal}{Phys. Rev.} \textbf{\bibinfo{volume}{D68}},
  \bibinfo{pages}{104012} (\bibinfo{year}{2003}{\natexlab{b}}),
  \eprint{gr-qc/03x1}.

\bibitem[{\citenamefont{Campbell et~al.}(1991)\citenamefont{Campbell, Duncan,
  Kaloper, and Olive}}]{CAMPBELL1991778}
\bibinfo{author}{\bibfnamefont{B.~A.} \bibnamefont{Campbell}},
  \bibinfo{author}{\bibfnamefont{M.}~\bibnamefont{Duncan}},
  \bibinfo{author}{\bibfnamefont{N.}~\bibnamefont{Kaloper}}, \bibnamefont{and}
  \bibinfo{author}{\bibfnamefont{K.~A.} \bibnamefont{Olive}},
  \bibinfo{journal}{Nuclear Physics B} \textbf{\bibinfo{volume}{351}},
  \bibinfo{pages}{778} (\bibinfo{year}{1991}), ISSN \bibinfo{issn}{0550-3213}.

\bibitem[{\citenamefont{Guarrera and Hariton}(2007)}]{Guarrera:2007tu}
\bibinfo{author}{\bibfnamefont{D.}~\bibnamefont{Guarrera}} \bibnamefont{and}
  \bibinfo{author}{\bibfnamefont{A.~J.} \bibnamefont{Hariton}}
  (\bibinfo{year}{2007}), \eprint{gr-qc/0702029}.

\bibitem[{\citenamefont{Grumiller and Yunes}(2008)}]{Grumiller:2007rv}
\bibinfo{author}{\bibfnamefont{D.}~\bibnamefont{Grumiller}} \bibnamefont{and}
  \bibinfo{author}{\bibfnamefont{N.}~\bibnamefont{Yunes}},
  \bibinfo{journal}{Phys. Rev.} \textbf{\bibinfo{volume}{D77}},
  \bibinfo{pages}{044015} (\bibinfo{year}{2008}), \eprint{0711.1868}.

\bibitem[{\citenamefont{Yunes and Pretorius}(2009)}]{Yunes:2009hc}
\bibinfo{author}{\bibfnamefont{N.}~\bibnamefont{Yunes}} \bibnamefont{and}
  \bibinfo{author}{\bibfnamefont{F.}~\bibnamefont{Pretorius}},
  \bibinfo{journal}{Phys. Rev. D} \textbf{\bibinfo{volume}{79}},
  \bibinfo{pages}{084043} (\bibinfo{year}{2009}), \eprint{0902.4669}.

\bibitem[{\citenamefont{Yagi et~al.}(2012)\citenamefont{Yagi, Yunes, and
  Tanaka}}]{Yagi:2012ya}
\bibinfo{author}{\bibfnamefont{K.}~\bibnamefont{Yagi}},
  \bibinfo{author}{\bibfnamefont{N.}~\bibnamefont{Yunes}}, \bibnamefont{and}
  \bibinfo{author}{\bibfnamefont{T.}~\bibnamefont{Tanaka}},
  \bibinfo{journal}{Phys. Rev. D} \textbf{\bibinfo{volume}{86}},
  \bibinfo{pages}{044037} (\bibinfo{year}{2012}), \bibinfo{note}{[Erratum:
  Phys.Rev.D 89, 049902 (2014)]}, \eprint{1206.6130}.

\bibitem[{\citenamefont{Maselli et~al.}(2017)\citenamefont{Maselli, Pani,
  Cotesta, Gualtieri, Ferrari, and Stella}}]{Maselli:2017kic}
\bibinfo{author}{\bibfnamefont{A.}~\bibnamefont{Maselli}},
  \bibinfo{author}{\bibfnamefont{P.}~\bibnamefont{Pani}},
  \bibinfo{author}{\bibfnamefont{R.}~\bibnamefont{Cotesta}},
  \bibinfo{author}{\bibfnamefont{L.}~\bibnamefont{Gualtieri}},
  \bibinfo{author}{\bibfnamefont{V.}~\bibnamefont{Ferrari}}, \bibnamefont{and}
  \bibinfo{author}{\bibfnamefont{L.}~\bibnamefont{Stella}},
  \bibinfo{journal}{Astrophys. J.} \textbf{\bibinfo{volume}{843}},
  \bibinfo{pages}{25} (\bibinfo{year}{2017}), \eprint{1703.01472}.

\bibitem[{\citenamefont{Penrose}(1965)}]{Penrose:1964wq}
\bibinfo{author}{\bibfnamefont{R.}~\bibnamefont{Penrose}},
  \bibinfo{journal}{Phys. Rev. Lett.} \textbf{\bibinfo{volume}{14}},
  \bibinfo{pages}{57} (\bibinfo{year}{1965}).

\bibitem[{\citenamefont{Hawking and Penrose}(1970)}]{Hawking:1969sw}
\bibinfo{author}{\bibfnamefont{S.~W.} \bibnamefont{Hawking}} \bibnamefont{and}
  \bibinfo{author}{\bibfnamefont{R.}~\bibnamefont{Penrose}},
  \bibinfo{journal}{Proc. Roy. Soc. Lond. A} \textbf{\bibinfo{volume}{314}},
  \bibinfo{pages}{529} (\bibinfo{year}{1970}).

\bibitem[{\citenamefont{Delbourgo and Salam}(1972)}]{Delbourgo:1972xb}
\bibinfo{author}{\bibfnamefont{R.}~\bibnamefont{Delbourgo}} \bibnamefont{and}
  \bibinfo{author}{\bibfnamefont{A.}~\bibnamefont{Salam}},
  \bibinfo{journal}{Phys. Lett. B} \textbf{\bibinfo{volume}{40}},
  \bibinfo{pages}{381} (\bibinfo{year}{1972}).

\bibitem[{\citenamefont{Alexander}(2005)}]{Alexander:2005vb}
\bibinfo{author}{\bibfnamefont{S.}~\bibnamefont{Alexander}},
  \bibinfo{journal}{Phys. Lett. B} \textbf{\bibinfo{volume}{629}},
  \bibinfo{pages}{53} (\bibinfo{year}{2005}), \eprint{hep-th/0503146}.

\bibitem[{\citenamefont{Silva et~al.}(2020)\citenamefont{Silva, Holgado,
  C\'ardenas-Avenda\~no, and Yunes}}]{Silva:2020acr}
\bibinfo{author}{\bibfnamefont{H.~O.} \bibnamefont{Silva}},
  \bibinfo{author}{\bibfnamefont{A.~M.} \bibnamefont{Holgado}},
  \bibinfo{author}{\bibfnamefont{A.}~\bibnamefont{C\'ardenas-Avenda\~no}},
  \bibnamefont{and} \bibinfo{author}{\bibfnamefont{N.}~\bibnamefont{Yunes}}
  (\bibinfo{year}{2020}), \eprint{2004.01253}.

\bibitem[{\citenamefont{Abbott et~al.}(2017)}]{TheLIGOScientific:2017qsa}
\bibinfo{author}{\bibfnamefont{B.~P.} \bibnamefont{Abbott}}
  \bibnamefont{et~al.} (\bibinfo{collaboration}{LIGO Scientific, Virgo}),
  \bibinfo{journal}{Phys. Rev. Lett.} \textbf{\bibinfo{volume}{119}},
  \bibinfo{pages}{161101} (\bibinfo{year}{2017}), \eprint{1710.05832}.

\bibitem[{\citenamefont{Riley et~al.}(2019)}]{Riley:2019yda}
\bibinfo{author}{\bibfnamefont{T.~E.} \bibnamefont{Riley}}
  \bibnamefont{et~al.}, \bibinfo{journal}{Astrophys. J. Lett.}
  \textbf{\bibinfo{volume}{887}}, \bibinfo{pages}{L21} (\bibinfo{year}{2019}),
  \eprint{1912.05702}.

\bibitem[{\citenamefont{Miller et~al.}(2019)}]{Miller:2019cac}
\bibinfo{author}{\bibfnamefont{M.~C.} \bibnamefont{Miller}}
  \bibnamefont{et~al.}, \bibinfo{journal}{Astrophys. J. Lett.}
  \textbf{\bibinfo{volume}{887}}, \bibinfo{pages}{L24} (\bibinfo{year}{2019}),
  \eprint{1912.05705}.

\bibitem[{\citenamefont{Yagi et~al.}(2013)\citenamefont{Yagi, Stein, Yunes, and
  Tanaka}}]{Yagi:2013mbt}
\bibinfo{author}{\bibfnamefont{K.}~\bibnamefont{Yagi}},
  \bibinfo{author}{\bibfnamefont{L.~C.} \bibnamefont{Stein}},
  \bibinfo{author}{\bibfnamefont{N.}~\bibnamefont{Yunes}}, \bibnamefont{and}
  \bibinfo{author}{\bibfnamefont{T.}~\bibnamefont{Tanaka}},
  \bibinfo{journal}{Phys. Rev. D} \textbf{\bibinfo{volume}{87}},
  \bibinfo{pages}{084058} (\bibinfo{year}{2013}), \bibinfo{note}{[Erratum:
  Phys.Rev.D 93, 089909 (2016)]}, \eprint{1302.1918}.

\bibitem[{\citenamefont{Kallosh et~al.}(1995)\citenamefont{Kallosh, Linde,
  Linde, and Susskind}}]{Kallosh:1995hi}
\bibinfo{author}{\bibfnamefont{R.}~\bibnamefont{Kallosh}},
  \bibinfo{author}{\bibfnamefont{A.~D.} \bibnamefont{Linde}},
  \bibinfo{author}{\bibfnamefont{D.~A.} \bibnamefont{Linde}}, \bibnamefont{and}
  \bibinfo{author}{\bibfnamefont{L.}~\bibnamefont{Susskind}},
  \bibinfo{journal}{Phys. Rev. D} \textbf{\bibinfo{volume}{52}},
  \bibinfo{pages}{912} (\bibinfo{year}{1995}), \eprint{hep-th/9502069}.

\bibitem[{\citenamefont{Shiromizu and Tanabe}(2013)}]{Shiromizu:2013pna}
\bibinfo{author}{\bibfnamefont{T.}~\bibnamefont{Shiromizu}} \bibnamefont{and}
  \bibinfo{author}{\bibfnamefont{K.}~\bibnamefont{Tanabe}},
  \bibinfo{journal}{Phys. Rev. D} \textbf{\bibinfo{volume}{87}},
  \bibinfo{pages}{081504} (\bibinfo{year}{2013}), \eprint{1303.6056}.

\bibitem[{grt()}]{grtensor}
\emph{\bibinfo{title}{{GRTensorII}}}, \bibinfo{note}{this is a package which
  runs within Maple but distinct from packages distributed with Maple. It is
  distributed freely on the World-Wide-Web from the address: {\tt
  http://grtensor.org}}.

\bibitem[{\citenamefont{Delsate et~al.}(2018)\citenamefont{Delsate, Herdeiro,
  and Radu}}]{Delsate:2018ome}
\bibinfo{author}{\bibfnamefont{T.}~\bibnamefont{Delsate}},
  \bibinfo{author}{\bibfnamefont{C.}~\bibnamefont{Herdeiro}}, \bibnamefont{and}
  \bibinfo{author}{\bibfnamefont{E.}~\bibnamefont{Radu}},
  \bibinfo{journal}{Phys. Lett. B} \textbf{\bibinfo{volume}{787}},
  \bibinfo{pages}{8} (\bibinfo{year}{2018}), \eprint{1806.06700}.

\bibitem[{\citenamefont{Sullivan et~al.}(2020)\citenamefont{Sullivan, Yunes,
  and Sotiriou}}]{Sullivan:2020zpf}
\bibinfo{author}{\bibfnamefont{A.}~\bibnamefont{Sullivan}},
  \bibinfo{author}{\bibfnamefont{N.}~\bibnamefont{Yunes}}, \bibnamefont{and}
  \bibinfo{author}{\bibfnamefont{T.~P.} \bibnamefont{Sotiriou}}
  (\bibinfo{year}{2020}), \eprint{2009.10614}.

\bibitem[{\citenamefont{Bah and Heidmann}(2020)}]{Bah}
\bibinfo{author}{\bibfnamefont{I.}~\bibnamefont{Bah}} \bibnamefont{and}
  \bibinfo{author}{\bibfnamefont{P.}~\bibnamefont{Heidmann}}
  (\bibinfo{year}{2020}), \eprint{2011.08851}.

\bibitem[{\citenamefont{{Raychaudhuri}}(1955)}]{1955PhRv...98.1123R}
\bibinfo{author}{\bibfnamefont{A.}~\bibnamefont{{Raychaudhuri}}},
  \bibinfo{journal}{Physical Review} \textbf{\bibinfo{volume}{98}},
  \bibinfo{pages}{1123} (\bibinfo{year}{1955}).

\bibitem[{\citenamefont{Borde}(1987)}]{Borde_1987}
\bibinfo{author}{\bibfnamefont{A.}~\bibnamefont{Borde}},
  \bibinfo{journal}{Classical and Quantum Gravity}
  \textbf{\bibinfo{volume}{4}}, \bibinfo{pages}{343} (\bibinfo{year}{1987}),
  \urlprefix\url{https://doi.org/10.1088/0264-9381/4/2/015}.

\bibitem[{\citenamefont{Alexander
  et~al.}(2021{\natexlab{a}})\citenamefont{Alexander, Jenks, and Yunes}}]{AJY}
\bibinfo{author}{\bibfnamefont{S.}~\bibnamefont{Alexander}},
  \bibinfo{author}{\bibfnamefont{L.}~\bibnamefont{Jenks}}, \bibnamefont{and}
  \bibinfo{author}{\bibfnamefont{N.}~\bibnamefont{Yunes}},
  \bibinfo{journal}{\textit{To Appear,}}  (\bibinfo{year}{2021}{\natexlab{a}}).

\bibitem[{\citenamefont{Alexander
  et~al.}(2021{\natexlab{b}})\citenamefont{Alexander, Gabadadze, Jenks, and
  Yunes}}]{AJGY}
\bibinfo{author}{\bibfnamefont{S.}~\bibnamefont{Alexander}},
  \bibinfo{author}{\bibfnamefont{G.}~\bibnamefont{Gabadadze}},
  \bibinfo{author}{\bibfnamefont{L.}~\bibnamefont{Jenks}}, \bibnamefont{and}
  \bibinfo{author}{\bibfnamefont{N.}~\bibnamefont{Yunes}},
  \bibinfo{journal}{\textit{To Appear,}}  (\bibinfo{year}{2021}{\natexlab{b}}).

\end{thebibliography}
\end{document}